\def\mode{1}
\if 0\mode
\documentclass[12pt,onecolumn,draftclsnofoot,letterpaper]{IEEEtran}
\let\chapter\section
\else
\documentclass[journal]{IEEEtran}
\fi
\usepackage{graphicx}
\usepackage{amsmath}
\usepackage{amssymb}
\usepackage{multirow}
\usepackage{tabularx}
\usepackage{listliketab}
\usepackage{cite}
\usepackage{url}
\usepackage{color}
\usepackage{amsthm}

\usepackage{algorithm}
\usepackage{algpseudocode}
\usepackage{pifont}
\usepackage{bm}
\newtheorem*{theorem*}{Theorem}
\usepackage{enumitem}
\newcommand{\mat}{\mathbf}
\renewcommand{\t}[1]{\text{#1}}
\renewcommand{\v}[1]{\mathbf{\bar{#1}}}
\newcommand{\C}[1]{\mathcal{#1}}
\newcommand{\bb}[1]{\mathbb{#1}}

\newcommand{\I}{I_{\rm{H}}}

\ifCLASSOPTIONcompsoc
\usepackage[caption=false,font=normalsize,labelfont=sf,textfont=sf]{subfig}
\else
\usepackage[subrefformat=parens,labelformat=parens,caption=false,font=footnotesize]{subfig}
\fi

\usepackage{array}
\newcolumntype{L}[1]{>{\raggedright\let\newline\\\arraybackslash\hspace{0pt}}m{#1}}
\newcolumntype{C}[1]{>{\centering\let\newline\\\arraybackslash\hspace{0pt}}m{#1}}
\newcolumntype{R}[1]{>{\raggedleft\let\newline\\\arraybackslash\hspace{0pt}}m{#1}}
\newcommand{\st}{\text{s.t.\hspace{5mm}}}
\begin{document}

\title{Harvest-or-Transmit Policy for Cognitive Radio Networks: A Learning Theoretic Approach}

\author{Kalpant~Pathak$^*$,~\IEEEmembership{Student Member,~IEEE,}
	and~Adrish~Banerjee$^{\dagger}$,~\IEEEmembership{Senior~Member,~IEEE}
\thanks{$^*$The author was with Indian Institute of Technology Kanpur. He is now with Qualcomm Corporate R\&D Bangalore, Karnataka, 560066, INDIA. e-mail: kalppath@qti.qualcomm.com.\newline\indent$^{\dagger}$The author is with the Department of Electrical Engineering, Indian Institute of Technology Kanpur, Uttar Pradesh, 208016, INDIA. e-mail: adrish@iitk.ac.in.\newline
\indent Part of this work is accepted in the 2018 International Conference on Signal Processing and Communications (SPCOM'18)\cite{my_spcom_18}.}}

\maketitle

\begin{abstract}
We consider an underlay cognitive radio network where the secondary user (SU) harvests energy from the environment. We consider a slotted-mode of operation where each slot of SU is used for either energy harvesting or data transmission. Considering block fading with memory, we model the energy arrival and fading processes as a stationary Markov process of first order. We propose a \textit{harvest-or-transmit} policy for the SU along with optimal transmit powers that maximize its expected throughput under three different settings. First, we consider a learning-theoretic approach where we do not assume any apriori knowledge about the underlying Markov processes. In this case, we obtain an online policy using \textit{Q-learning}. Then, we assume that the full statistical knowledge of the governing Markov process is known apriori. Under this assumption, we obtain an optimal online policy using infinite horizon \textit{stochastic dynamic programming}. Finally, we obtain an optimal offline policy using the \textit{generalized Benders decomposition} algorithm. The offline policy assumes that for a given time deadline, the energy arrivals and channel states are known in advance at all the transmitters. Finally, we compare all policies and study the effects of various system parameters on the system performance.

\end{abstract}


\IEEEpeerreviewmaketitle

\section{Introduction}
\IEEEPARstart{T}{he} performance of a wireless communication network critically depends on the spectrum and energy availability at communication nodes. In the conventional spectrum allocation schemes, the spectrum is mainly owned by the primary users (PUs), and the secondary users (SUs) do not have access to the licensed spectrum. Cognitive radio network (CRN) has emanated as a possible solution to the problem of insufficient availability of spectrum \cite{CR_simon_haykin}. In CRNs, SUs are aware of their environment and opportunistically access the licensed spectrum such that the quality of service requirements of PUs are maintained. 


Applications such as wireless sensor networks may require sensor nodes to be placed in secluded areas, which makes their regular maintenance a strenuous task. Also, with the emergence of Internet-of-Things (IoT) where billions of devices wish to share information with each other, energy availability at such communicating nodes is a challenging issue. In such scenarios, harvesting energy from the environmental sources can ensure perennial operation of such communicating devices \cite{EH_survey_1,EH_survey_2}. Together with the CRN, energy harvesting (EH) can be used to mitigate the issue of spectrum scarcity and energy availability simultaneously. The major challenge in incorporating energy harvesting in CRNs is the random amplitudes and arrivals of the energy. Due to which, designing optimal transmission policies for the communicating node that maximize the network performance becomes essential. The framework of designing a transmission policy can be classified into: online framework, and offline framework. In online policies, a transmitter has either statistical or no knowledge of future channel coefficients and harvested energies. On the other hand, in the offline policies, we assume that the future channel coefficients and harvested energies are completely known at the transmitter.

\subsection{Background Works}
\label{subsec:background}
Optimal transmission policies for underlay EH-CRNs have been addressed well in the literature \cite{underlay_1,underlay_2,underlay_SWIPT_MIMO,underlay_3,my_spcom,my_globecom,underlay_pwr_time_Xu,underlay_multihop_Xu,underlay_multihop,underlay_hybrid}. In \cite{underlay_1}, the authors obtained an optimal offline power control policy for EH-secondary transmitter (ST) using  geometric water-filling with peak power constraints that maximized ST's throughput. In \cite{underlay_2}, the authors considered energy cooperation between the EH-ST and EH-primary transmitter (PT) and obtained myopic and offline cooperation protocols in single and multi-slot settings, respectively, that maximized the throughput of SU. In \cite{underlay_SWIPT_MIMO}, the authors designed a robust transceiver for the simultaneous wireless information and power transfer (SWIPT) based multiple-input multiple-output (MIMO) underlay CRN. The work in \cite{underlay_3}, \cite{my_spcom}, and \cite{my_globecom} considered an underlay EH-CRN where the ST harvested energy from PT's interference in \cite{underlay_3} and \cite{my_spcom}, and from other environmental sources in \cite{my_globecom}. The authors obtained a myopic transmission policy in \cite{underlay_3}, an optimal offline policy in \cite{my_spcom} and a robust online policy in \cite{my_globecom}. In \cite{underlay_pwr_time_Xu}, \cite{underlay_multihop_Xu}, and \cite{underlay_multihop}, authors considered a multihop underlay EH-CRN network where the energy harvesting ST communicated with the secondary receiver (SR) over multiple hops using the time division multiple access (TDMA) protocol. Under the interference constraint of PU, the authors obtained optimal time-sharing among SUs that maximized the sum-rate of SU in \cite{underlay_pwr_time_Xu}, the end-to-end throughput in \cite{underlay_multihop_Xu}, and minimized the outage probability in \cite{underlay_multihop}. In \cite{underlay_hybrid}, authors considered a hybrid overlay-underlay EH-CRN where the SU harvested energy from the transmission of PU as well as ambient sources. The authors employed the partially observable Markov decision process (POMDP) framework and derived an energy threshold (to decide the transmission mode) to obtain an online policy for SU that maximized its throughput. In addition to the works on throughput maximization considering energy arrival constraints, \cite{delay_constraint_1,delay_constraint_2} considered delay constrained energy harvesting networks where both data and energy arrivals are sporadic in nature. The authors in \cite{delay_constraint_1} considered a point-to-point link whereas the authors in \cite{delay_constraint_2} considered a CRN and obtained optimal energy scheduling in both scenarios.

The aforementioned works on optimal transmission policies in underlay EH-CRNs focused mainly on myopic \cite{underlay_2,underlay_SWIPT_MIMO,underlay_3,underlay_pwr_time_Xu,underlay_multihop_Xu,underlay_multihop}, offline \cite{underlay_1,underlay_2,my_spcom} and online policies \cite{my_globecom,underlay_hybrid,delay_constraint_2}. The myopic policies aim to maximize the immediate throughput and do not care about the past and future channel conditions and energy arrivals, which is suboptimal in multi-slot scenarios. The offline policies, on the other hand, assume that the channel conditions and energy arrivals are known non-causally at the transmitters. This information allows the transmitters to adapt their transmission powers according to channel conditions and energy availability, which results in higher achievable throughput \cite{my_spcom}. However, the non-causal knowledge of channel gains and energy arrivals at the transmitters may be an unreasonable assumption if the environment is highly dynamic. These dynamics can be incorporated in online policies that assume only the statistical knowledge about the channel conditions and energy arrivals and therefore, are more practical in terms of implementation. However, complete statistical knowledge of energy arrivals and channel conditions is still a strong assumption and may not be valid in many practical scenarios.

In scenarios where the transmitters do not have any statistical knowledge of the channel conditions and energy arrivals, the optimal transmission policies can be obtained using a reinforcement learning (RL) based approach. The framework of RL based algorithms requires the optimization problem to be formulated as a Markov decision process (MDP) \cite{DP_powell,MDP_puterman}. In literature, there are few works on the application of RL based algorithms in designing the optimal transmission policies in point-to-point EH networks \cite{learning_dohler,learning_point_to_point,learning_Hsu,learning_Prabuchandran}. The works in \cite{RL_CRN_Yau,RL_CRN_Raj,RL_CRN_Jang,RL_CRN_Arunthavanathan,RL_CRN_Lunden,RL_CRN_Mendes,RL_CRN_Valehi,RL_EH_CRN_Shafie} considered the application of RL in CRNs. The authors in \cite{RL_CRN_Yau} proposed various applications of RL algorithms in CRNs. Conventional interweave CRNs have been studied in \cite{RL_CRN_Raj,RL_CRN_Jang,RL_CRN_Arunthavanathan,RL_CRN_Lunden,RL_CRN_Mendes,RL_CRN_Valehi} where the authors proposed RL based spectrum sensing policies. In \cite{RL_EH_CRN_Shafie}, the authors considered an overlay EH-CRN where the EH-SU helped EH-PU deliver its data. The authors obtained RL based transmission and cooperation strategy for SU that maximized its throughput.

\subsection{Motivation and Contribution}
\label{subsec:motivation}
In scenarios where the wireless environment is highly dynamic, it is difficult to predict the future wireless conditions with significant accuracy and obtain the benchmark offline resource allocation. In such cases, dynamic programming and RL based policies can be very beneficial in optimally allocating resources among the nodes. In some cases, even the statistical knowledge about the channel gains and energy arrival processes is not completely known. This motivated us to employ the RL based algorithms in designing optimal transmission policies for EH-CRN aiming to maximize the spectral and energy efficiency of the secondary network simultaneously.

We consider an underlay EH-CRN where the ST scavenges energy from the environment and communicates with the SR in slotted fashion. The transmit power of the PT is assumed constant in all slots. Different from the system model proposed in \cite{underlay_2,my_spcom,my_globecom} where each slot was shared between \textit{energy harvesting} and \textit{data transmission} (time-sharing model), in our model, each slot is either used for energy harvesting or data transmission. Considering harvest-or-transmit model over the time-sharing model reduces the size of action space which reduces the complexity of dynamic programming algorithm for the online policy \cite{learning_book_haykin}. To employ dynamic programming algorithms for solving the discrete time MDPs, we need to quantize the action space (which might be transmit power and time-sharing parameter). For the time-sharing based models considered in \cite{underlay_2,my_spcom,my_globecom}, if time-sharing parameter and transmit power can take values from finite sets with $N_T$ and $N_P$ number of elements, respectively, the total size of the action set would be $N_T \times N_P$. On the other hand, in the harvest-or-transmit model, the size of action set would be $N_P + 1$. Also, we assume that in each slot, SU may terminate its operation with a non-zero probability. We model the energy arrival and fading process as a first-order stationary Markov process as in \cite{learning_dohler,learning_point_to_point,learning_Hsu,learning_Prabuchandran}, which enables us to employ tools from dynamic programming and RL.

Our aim is to obtain an optimal harvest-or-transmit policy where, in each slot, the ST optimally decides whether to harvest or transmit depending on the channel conditions and energy availability. The ST chooses its transmit power in each slot such that it maximizes its sum-rate and keeps the worst-case interference at the primary receiver (PR) below a threshold. 

\begin{table*}[!ht]
	\centering
	\caption{Scenarios}
	\begin{tabular}{|C{2.5cm}|C{8cm}|C{4cm}|}
		\hline
		\textbf{Scenario} & \textbf{Assumptions} & \textbf{Solution}\\
		\hline
		Offline & Non-causal knowledge of energy arrivals and channel coefficients & GBD algorithm\\
		\hline
		Online & Transition probabilities of energy and channel states are known& Policy Iteration algorithm\\
		\hline
		Learning Theoretic & Only instantaneous values of harvested energy and channel gains are known& Q-learning algorithm\\
		\hline
		Myopic \cite{underlay_3} & Only instantaneous values of harvested energy and channel gains are known & CVX \cite{cvx}\\
		\hline
	\end{tabular}
	\label{table:scenarios}
\end{table*}
\begin{table*}[!t]
	\centering
	\caption{Notation}
	\begin{tabular}{|C{2.5cm}|L{12.5cm}|}
		\hline
		$\eta$ & Energy harvesting efficiency\\
		\hline
		$\gamma$ & Probability of ST's termination of operation\\
		\hline
		$P_s^{[i]}$ & Transmit power of ST in $i$th slot\\
		\hline
		$I_{\rm H}^{[i]}$ & Indicator variable denoting harvest or transmit decision\\
		\hline
		$B_0$ & Initial energy in the battery at ST\\
		\hline
		$B_{\rm max}$ & Battery capacity of ST\\
		\hline
		$\tau$ & Slot length in seconds\\
		\hline
		$P_{\rm int}$ & Worst-case interference constraint of PR\\
		\hline
		$\sigma_n^2$ & Variance of noise at secondary receiver (SR)\\
		\hline
		$P_p$ & Transmit power of primary transmitter (PT)\\
		\hline
		$h_{ss}^{[i]},h_{sp}^{[i]},h_{pp}^{[i]},h_{ps}^{[i]}$ & Channel power gain of ST-SR, ST-PR, PT-PR, and PT-SR links in the $i$th slot, respectively\\
		\hline
		$E_{\rm H}^{[i]}$ & Energy available for harvesting in $i$th slot\\
		\hline
		$P_{\rm max}$ & Maximum transmit power of ST\\
		\hline
		$\mathcal{S}$ & Set of all possible system states\\
		\hline
		$\mathcal{A}$ & Set of all possible actions\\
		\hline
		$N_S$ & Number of states (cardinality of $\mathcal{S}$)\\
		\hline
		$N_A$ & Number of possible actions (cardinality of $\mathcal{A}$)\\
		\hline
		$\mathbf{R}$ & Reward matrix of dimension $N_S\times N_S\times N_A$\\
		\hline
		$\mathbf{T}$  & State transition probability matrix of dimension $N_S\times N_S\times N_A$\\
		\hline
		$\pi$ & Policy\\
		\hline
	\end{tabular}
	\label{table:notation}
\end{table*}
The contributions of this paper are as follows:
\begin{enumerate}
	\item We first assume that the ST has full statistical knowledge about the underlying energy arrival and fading processes. Under this assumption, we formulate the sum-rate maximization problem as a Markov decision process (MDP) \cite{DP_powell}, and obtain the online harvest-or-transmit policy using the \textit{policy iteration} algorithm \cite{DP_powell}.
	\item We then consider a case where the ST has no statistical knowledge about that channel fading and energy arrival processes. However, it can observe the instantaneous values of channel coefficients and harvested energy in each slot. In this scenario, we employ tools from RL and obtain the optimal harvest-or-transmit policy using \textit{Q-learning} algorithm \cite{DP_powell}.
	\item Next, we consider the offline optimization framework where the ST has non-causal knowledge of all channel coefficients and energy arrivals. Under the offline framework, the optimization problem is a non-convex mixed integer non-linear program (MINLP) with coupled variables \cite{cvx_book}. We decouple the variables and reduce it to a convex MINLP \cite{convex_MINLP}, and employ the generalized Benders decomposition (GBD) algorithm \cite{GBD} to solve it efficiently.
	\item  Finally, we compare the performance of all three policies and show that the offline policy acts as a benchmark for online and learning-theoretic policies. We show that the learning theoretic policy performs well for a large number of learning iterations. We also analyze the effects of the number of learning iterations, action-selection probability, PU's transmit power, and maximum transmit power of SU on the proposed policies. In addition, we compare proposed policies with the myopic policy proposed in the literature \cite{underlay_3} and show that our policies outperform the myopic policy in terms of achievable throughput.
\end{enumerate}
Different scenarios along with their assumptions and proposed solutions are summarized in Table \ref{table:scenarios}.

\subsection{Paper Organization and Notation}
The organization of the paper is as follows. The system model and the problem formulations are presented in Section \ref{sec:system_model} and Section \ref{sec:problem}, respectively. Section \ref{sec:MDP} discusses the online harvest-or-transmit policy using the MDP. Section \ref{sec:learning} presents the harvest-or-transmit policy using Q-learning, and Section \ref{sec:offline} presents the optimal offline policy using the GBD algorithm. The simulation results are discussed in Section \ref{sec:result}, and finally, we conclude in Section \ref{sec:conclusion}.

\textit{Notation:} A bold symbol with a bar (\textit{e.g.,} $\v{a}$ or $\v{\pmb{\theta}}$) represents an $N$-dimensional vector, without a bar (\textit{e.g.,}$\mathbf{M}$ or $\pmb{\theta}$) represents a matrix and $x^{[i]}$ represents the $i$th element of the vector $\v{x}$. $\v{x}\preceq\v{0}$ represents $x^{[i]}\leq 0,\forall i$, $[u]^+$ represents $\max\{0,u\}$, $\mathbb{E}[\cdot]$ represents the expectation operator, and $\mathbb{P}(\cdot)$ denotes the probability. The calligraphic symbols (e.g., $\mathcal{A}$) represents a set. and $\mathbb{R}_+^m$ represents a set of $m$-dimensional positive-real valued vectors. Other notations used in the paper are described in Table \ref{table:notation}.

\section{System Model}
\label{sec:system_model}
In this section, we present our system model. The Section \ref{subsec:network} presents the network model of underlay EH-CRN operating in the slotted mode, Sections \ref{subsec:channel_model} and \ref{subsec:energy_process} present the modeling of fading and energy harvesting processes, respectively, and Section \ref{subsec:battery_dynamics} presents the battery dynamics at the ST.

\subsection{Cognitive Radio Network}
\label{subsec:network}
\begin{figure}[!ht]
\centering
\includegraphics[width=\linewidth]{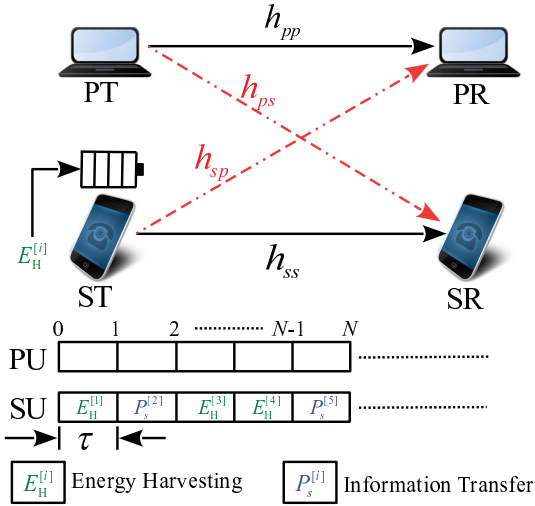}
\caption{An underlay EH-CRN.}
\label{fig:system_model}
\end{figure}
Fig. \ref{fig:system_model} shows an underlay EH-CRN where a PU and an SU operate in slotted fashion with a slot length of $\tau$ seconds. We assume the slot length to be 1 second so that we can use power and energy interchangeably. However, the proposed policies can be modified for any value of $\tau$ without loss of generality. The PT has a reliable energy source, while the ST scavenges energy from the environment in a half-duplex fashion. The transmit power of PT $P_p$ is constant in all the slots, whereas in $i$th slot, the ST decides whether to harvest energy from the environment with efficiency $0<\eta\leq1$ or transmit its data with power $P_s^{[i]}$. We represent the vector containing the transmit powers of the ST with $\v{p}_s$ such that $\v{p}_s[i]=P_s^{[i]}$, where $\v{p}_s[i]$ is the $i$th element of vector $\v{p}_s$. In any slot, the ST may terminate its operation with probability $(1-\gamma)$, $0<\gamma<1$, and is allowed to cause a worst-case interference $P_{\rm int}$ to the PR. We assume that the battery at the ST has an initial energy $B_0$ and capacity $B_{\rm{max}}$.

\subsection{Channel Model}
\label{subsec:channel_model}
We consider block fading with memory where the power gains of the channel links are modeled using the stationary Markov process of first order as in \cite{learning_dohler}. The channel power gains of ST-SR, ST-PR, PT-PR and PT-SR links are represented by $\v{h}_{ss}$, $\v{h}_{sp}$, $\v{h}_{pp}$, and $\v{h}_{ps}$, respectively. In the $i$th slot, the channel power gains ($h_{ss}^{[i]},h_{sp}^{[i]},h_{ps}^{[i]}$ and $h_{pp}^{[i]}$) can take any value from a finite set $\mathcal{H}\triangleq\{h_1,h_2,\ldots,h_{M_C}\}$ independently, where $M_C$ is the number of channel states. We assume that within each slot, the channel power gains remain constant and change from one slot to another with a transition probability $\mathbb{P}(h_i,h_j)$, $i,j\in\{1,\ldots,M_C\}$. Since the channel power gains follow a first-order stationary Markov process, for any slot $1\leq i\leq N$ we have
\begin{align*}
\mathbb{P}\left(h^\mathcal{C}_i|h^\mathcal{C}_{i-1},\ldots,h^\mathcal{C}_{1}\right)=\mathbb{P}\left(h^\mathcal{C}_i|h^\mathcal{C}_{i-1}\right),\; \forall i,
\end{align*}
where $\mathcal{C}=\{\text{PT-PR, PT-SR, ST-PR, ST-SR}\}$ represents a channel link between any transmitter-receiver pair.

In a transmission slot, say slot $i$, the ST transmits a signal $s^{[i]}$ of power $P_s^{[i]}=\mathbb{E}\left[|s^{[i]}|^2\right]$, and the PT transmits a signal $q^{[i]}$ with a constant power $P_p = \mathbb{E}\left[|q^{[i]}|^2\right]$. The received signal $r_i$ at the SR is given as
\begin{align*}
r^{[i]} = \sqrt{h_{ss}^{[i]}}s^{[i]}+\sqrt{h_{ps}^{[i]}}q^{[i]}+n^{[i]},
\end{align*}
where $n^{[i]}$ is the additive white Gaussian noise (AWGN) at the SR with mean $0$ and variance $\sigma_n^2$. The instantaneous achievable rate of ST in the $i$th transmission slot (in bpcu) is given as \cite{info_theory_book}
\begin{align}
R_{\t{ST}}^{[i]} = \gamma^i\log_2\left(1+\frac{h_{ss}^{[i]}P_s^{[i]}}{\sigma_n^2+h_{ps}^{[i]}P_p}\right).
\label{eq:instant_rate}
\end{align}

\subsection{Energy Harvesting Process}
\label{subsec:energy_process}
In an energy harvesting wireless network, the energy arrival time and amounts are random. We model the energy arrival process as a correlated time process following a stationary Markov process of first order and $M_E$ number of states as in \cite{learning_dohler}. In the $i$th slot, $E_{\rm{H}}^{[i]}$ Joules of energy arrives at the ST. However, the ST chooses either to harvest and store this energy in its finite-capacity battery or to transmit its data depending on channel conditions and the energy available in the battery. In each slot, the $E_{\rm{H}}^{[i]}$ can take value from a finite set $\mathcal{E}\triangleq\{e_1,e_2,\ldots e_{M_E}\}$. Similar to channel power gains, the $E_{\rm{H}}^{[i]}$ is constant within a time slot and changes from one slot to another with a transition probability $\mathbb{P}(e_i,e_j)$, $i,j\in\{1,\ldots,M_E\}$. Since $E_{\rm{H}}^{[i]},\,\forall i$ follows a Markov process, we have
\begin{align*}
\mathbb{P}\left(E_{\rm{H}}^{[i]}| E_{\rm{H}}^{[i-1]},\ldots,E_{\rm{H}}^{[1]}\right)=\mathbb{P}\left(E_{\rm{H}}^{[i]}|E_{\rm{H}}^{[i-1]}\right),\; \forall i.
\end{align*}
\subsection{Battery Dynamics}
\label{subsec:battery_dynamics}
The battery at the ST has a finite capacity of $B_{\rm{max}}$. In each slot, energy is either stored in the battery or drawn from it depending on the decision made by ST. At the beginning of each slot, the available energy in the battery depends on the energy harvested or consumed in the previous slot. Thus, the battery state also follows the first-order stationary Markov process.

To characterize the \textit{harvest-or-transmit} decision of the ST, we define an indicator variable $I^{[i]}_{\rm{H}}$ such that
\begin{align*}
I^{[i]}_{\rm{H}}=\left\{
\begin{array}{ll}
1, & \text{if ST harvests energy in } i^{th} \text{ slot},\\
0, & \text{if ST transmits in } i^{th}\text{ slot}.
\end{array}
\right.
\end{align*}
Whenever $I^{[i]}_{\rm{H}}=1$, the ST stores $E^{[i]}_{\rm{H}}$ amount of harvested energy to the battery and when $I^{[i]}_{\rm{H}}=0$, ST draws $ P_s^{[i]}$ energy from the battery in the $i$th slot. We define a vector $\v{i}_{\rm H}$ such that the $i$th element of the vector $\v{i}_{\rm H}[i]=I_{\rm H}^{[i]}$. If $B_i$ denote battery state at the beginning of the $i$th slot, then at the beginning of the ($i+1$)th slot, the battery state is given as
\begin{align}
B_{i+1} = &\min\left\{B_i+I_{\rm{H}}^{[i]}\eta E^{[i]}_{\rm{H}}-\left(1-I^{[i]}_{\rm{H}}\right) P_s^{[i]}, B_{\rm{max}}\right\},\label{eq:battery}
\end{align} 
where $0<\eta<1$ is the energy harvesting efficiency.
\subsubsection*{Energy Neutrality Constraint}
In networks with energy harvesting capabilities, the energy arrives during the data transmission which puts an extra constraint on the transmit power of a node, known as energy neutrality constraint. This constraint states that at any instant, the transmitter can use only that much energy which is available in the battery at that instant, \textit{i.e.,} for the $i$th slot we have
\begin{align}
\left(1-I^{[i]}_{\rm{H}}\right) P_s^{[i]} \leq B_i, \quad \forall i. \label{eq:power_battery}
\end{align}
Using \eqref{eq:battery}, the energy neutrality constraint in \eqref{eq:power_battery} can be rewritten non-recursively as
\begin{align}
\sum_{j=1}^i \left(1-I_{\rm{H}}^{[j]}\right) P_s^{[j]}&\leq B_0+ \eta \sum_{j=0}^{i}I^{[j]}_{\rm{H}}E_{\rm{H}}^{[j]},\;\forall i, \label{eq:energy_causality_1}\\
\sum_{j=l}^i \left(1-I_{\rm{H}}^{[j]}\right) P_s^{[j]}&\leq B_{\rm max}+\sum_{j=l}^i\eta I_{\rm H}^jE_{\rm H}^j,\; \forall i, l=1,\ldots,i. \label{eq:energy_causality_2}
\end{align}
\subsection{Worst-Case Interference at the PR}
\label{subsec:worst_case_interference}
We assume that the PU has a worst-case interference constraint $P_{\rm int}$ at the PR. In each slot, the ST chooses its transmit power such that the worst-case interference at the PR remains less than or equal to $P_{\rm int}$. Since channel power gains take values from a finite set $\mathcal{H}\triangleq\{h_1,h_2,\ldots,h_{M_C}\}$, the worst-case interference at the PR is given as
\begin{align}
P_{\rm int}=h_{\rm best}P_{\max}, \label{eq:worst_interference}
\end{align}
where $h_{\rm best}=\max\{h_1,h_2,\ldots,h_{M_C}\}$, and $P_{\max}$ is the transmit power constraint of the ST. From \eqref{eq:worst_interference} we have
\begin{align}
P_{\max}=\frac{P_{\rm int}}{h_{\rm best}}. \label{eq:max_ST_power}
\end{align}
In other words, to satisfy the worst-case interference constraint, we must have
\begin{align*}
P_s^{[i]}\leq P_{\max}.
\end{align*}
The MDP framework models a class of situations where the
agent/user has to take a decision whose outcomes are moderately random and moderately under the control of the decision-making agent \cite{DP_powell,MDP_puterman}. The system model discussed so far constitutes a discrete-time MDP with a finite number of states. An MDP is defined by a quadruplet $\langle\mathcal{S},\mathcal{A},\mathbb{P}_{a_i}(s_j,s_k),R_{a_i}(s_j,s_k) \rangle$, where $\mathcal{S}$ is the set of all possible states of the governing Markov process, $\mathcal{A}$ is the set of all possible actions the agent can take, $\mathbb{P}_{a_i}(s_j,s_k)$ denotes the state transition probability from state $s_j\in\mathcal{S}$ to $s_k\in\mathcal{S}$ when an action $a_i\in\mathcal{A}$ is taken, and $R_{a_i}(s_j,s_k)$ is the immediate reward the agent receives when it takes an action $a_i\in\mathcal{A}$ and its state changes from $s_j\in\mathcal{S}$ to $s_k\in\mathcal{S}$.

In the system model under consideration, the channel links that affect the rate achieved by the ST are PT-SR and ST-SR. Thus in the $i$th slot, the state of the governing Markov process, $s_i$ is comprised of four elements $s_i=\left\{h_{ps}^{[i]}, h_{ss}^{[i]},E^{[i]}_{\rm{H}},B_i\right\}$. Since we assumed that these states take values from finite sets, the possible number of states in set $\mathcal{S}$ will also be finite. The action taken by the ST in the $i$th slot is a vector comprised of two variables $a_i=\left[\begin{array}{cc}
I^{[i]}_{\rm{H}} & P_s^{[i]}\end{array}\right]^T$ with $I^{[i]}_{\rm{H}}\in\mathcal{I}\triangleq\{0,1\}$ and $P_s^{[i]}\in\mathcal{P}\triangleq\{0,P_1,P_2,\ldots, P_{\rm{max}}\}$, where $P_{\rm{max}}$ is the maximum transmit power constraint of the ST given in \eqref{eq:max_ST_power}. Note that when $\I^{[i]}=1$, $P_s^{[i]}=0$ and when $\I^{[i]}=0$, $P_s^{[i]}\in\mathcal{P}\backslash\{0\}$. Thus, the total number of possible actions $N_A$ is same as the cardinality of the set $\mathcal{P}$. We define a state transition matrix $\mat{T}$ of dimension $N_S\times N_S\times N_A$ such that $[\mat{T}]_{j,k,i}=\mathbb{P}_{a_i}(s_j,s_k)$, where $N_S$ is the number of possible states (total number of elements in set $\mathcal{S}$). The immediate reward $R_{a_i}(s_j,s_k)$ in our model is the instantaneous rate achieved by the ST in $i$th slot, which is given by \eqref{eq:instant_rate}.

\section{Problem Formulation}
\label{sec:problem}
The aim is to obtain a deterministic and stationary \textit{harvest-or-transmit} policy $\pi(\cdot):\mathcal{S}\mapsto \mathcal{A}$, that maximizes the expected sum-rate of ST subject to the energy neutrality and transmit power constraints before it terminates its operation. The optimization problem is given as
\begingroup
\allowdisplaybreaks
\begin{subequations}
\label{eq:online_orig}
\begin{align}
\max_{\left\{\I^{[i]},P_s^{[i]}\right\}_{i=1}^\infty}\quad& \lim_{N\to\infty}\mathbb{E}\left[ \sum_{i=1}^N\left(1-\I^{[i]}\right)R_{\rm ST}^{[i]}\right] \label{eq:online_orig_obj}\\
\t{s.t.\hspace{6mm}}\quad &\hspace{-2mm} \sum_{j=1}^i\left(1-I^{[j]}_{\rm H}\right) P_s^{[j]}\leq B_0+\eta\sum_{j=0}^{i}I^{[j]}_{\rm H}E_{\rm H}^{[j]},\;\forall i,\nonumber\\
&(\t{Energy neutrality constraint})\label{eq:online_orig_c1}\\
&\hspace{-2mm}\sum_{j=l}^i \left(1-I_{\rm{H}}^{[j]}\right) P_s^{[j]}\leq B_{\rm max}+\sum_{j=l}^i\eta I_{\rm H}^jE_{\rm H}^j,\nonumber\\
&\qquad\qquad\qquad\qquad \quad \forall i,\, l=1,\ldots,i,\label{eq:online_orig_c2}\\
& (\t{Battery capacity constraint}) \nonumber\\
&P_s^{[i]}\leq P_{\rm{max}},\;\; \forall i, \label{eq:online_orig_c3}\\
&(\t{Maximum transmit power constraint})\nonumber\\
& \I^{[i]}\in\mathcal{I},\;\;\;P_s^{[i]}\in\mathcal{P},\;\;\forall i, \label{eq:online_orig_c4}
\end{align}
\end{subequations}
\endgroup
where, the constraint \eqref{eq:online_orig_c2} is added due to the finite battery capacity and the expectation is taken with respect to all possible states. Note that $I_{\rm H}^{[0]}=E_{\rm H}^{[0]}=0$. Using the recursive relation of the battery state in \eqref{eq:battery} and the transmit power constraint in \eqref{eq:power_battery}, the optimization problem in \eqref{eq:online_orig} can be rewritten as
\begingroup
\allowdisplaybreaks
\begin{subequations}
\label{eq:online}
\begin{align}
\max_{\left\{\I^{[i]},P_s^{[i]}\right\}_{i=1}^\infty}\quad&\lim_{N\to\infty}\mathbb{E}\left[ \sum_{i=1}^N\left(1-\I^{[i]}\right)R_{\rm ST}^{[i]}\right] \label{eq:online_obj}\\
{\st\hspace{2mm}}\quad &\left(1-I_{\rm{H}}^{[i]}\right) P_s^{[i]} \leq B_i, \quad\forall i, \label{eq:online_c1}\\
& B_{i+1} = \min\{\Theta, B_{\rm{max}}\}, \quad\forall i, \label{eq:online_c2}\\
&  P_s^{[i]}\leq P_{\rm{max}},\quad\forall i,\label{eq:online_c3}\\
& \I^{[i]}\in\mathcal{I}=\{0,1\},\;\;\;P_s^{[i]}\in\mathcal{P}, \quad\forall i,\label{eq:online_c4}
\end{align}
\end{subequations}
\endgroup
where $\Theta:=B_i+I_{\rm{H}}^{[i]}\eta E^{[i]}_{\rm{H}}-\left(1-I^{[i]}_{\rm{H}}\right) P_s^{[i]}$. The next state of the battery $B_{i+1}$ can be uniquely determined for a given policy $\pi$ and current state $S_i$, whereas other components $\left(h_{ps}^{[i+1]},h_{ss}^{[i+1]},E^{[i+1]}_{\rm{H}}\right)$ can be determined probabilistically from the state transition matrix $\mat{T}$. Since the next state of the system can be determined solely by the current state-action pair, the system model follows the Markov property. This facilitates us to use tools from dynamic programming (DP) and reinforcement learning (RL) to obtain an online harvest-or-transmit policy.

We first define two functions, namely, the \textit{state-value function} and the \textit{action-value function} for a given policy $\pi$ and state $s_j$ \cite{DP_powell}. The state-value function $V^{\pi}(s_j)$, which is the reward of being in state $s_j$ is given as
\begin{align}
V^{\pi}(s_j)\triangleq C_{s_j}^{s_k}(\pi(s_j))+\gamma\sum_{\forall s_k\in\C{S}}\bb{P}_{\pi(s_j)}(s_j,s_k) V^{\pi}(s_k),\label{eq:state_value}
\end{align}
where $C_{s_j}^{s_k}(\pi(s_j))=\sum\limits_{\forall s_k\in\C{S}}\bb{P}_{\pi(s_j)}(s_j,s_k)R_{\pi(s_j)}(s_j,s_k)$ and $0<\gamma<1$ represent the immediate expected reward when an action $\pi(s_j)$ taken in state $s_j$ under policy $\pi$ and the state changes to $s_k$, and the discount factor, respectively. The action-value function $Q^\pi(s_j,a_i)$ represents the expected discounted sum-reward when an action $a_i$ is taken in state $s_j$. The system follows the policy $\pi$ after that. The action-value function is given as
\begin{align}
Q^\pi(s_j,a_i)\triangleq C_{s_j}^{s_k}(a_i)+\gamma\sum_{\forall s_k\in\C{S}}\bb{P}_{a_i}(s_j,s_k) V^{\pi}(s_k). \label{eq:action_value}
\end{align}
The state value function $V^\pi(s_j)$ characterizes the goodness of a policy $\pi$ is. We say a policy $\tilde{\pi}$ is a better policy than $\pi$ if $V^{\tilde{\pi}}(s_j)\geq V^{\pi}(s_j),\; \forall s_j\in\C{S}$, and a policy $\pi^*$ is optimal if $V^{\pi^*}(s_j)\geq V^{\pi}(s_j),\; \forall s_j\in\C{S}$ for all possible policies $\pi$. When ST follows an optimal policy $\pi^*$, we have
\begin{align}
V^{\pi^*}(s_j)=\max_{a_i\in\C{A}}\;\;Q^{\pi^*}(s_j,a_i). \label{eq:state_value_optimal}
\end{align}
This means that the optimal policy $\pi^*$ is \textit{greedy} with respect to $V^{\pi^*}(s_j)$.
Under the optimal policy, the action-value function is given as
\begin{align}
Q^{\pi^*}(s_j,a_i)=C_{s_j}^{s_k}(a_i)+\gamma\sum_{\forall s_k\in\C{S}}\bb{P}_{a_i}(s_j,s_k)\max_{a_i\in\C{A}}Q^{\pi^*}(s_k,a_i).\label{eq:action_value_optimal}
\end{align}
Eq. \eqref{eq:action_value_optimal} indicates that the action-value function under the optimal policy $\pi^*$ can be written as a sum of immediate expected reward $C_{s_j}^{s_k}(a_i)$ and the maximum value of action-value function in the next state $\max_{a_i\in\mathcal{A}}Q^{\pi^*}(s_k,a_i)$ \cite{DP_powell}.
\subsection*{Bellman's Optimality Criterion:}
The Bellman's optimality criterion \cite{DP_powell,MDP_puterman} says that irrespective of the first state and decision, an optimal policy must be followed starting from the state resulting from the first decision.

We consider three different frameworks to solve the expected sum-rate maximization problem in \eqref{eq:online} depending on the availability of information regarding system parameters. In a case where the ST knows the state transition matrix $\mat{T}$ and the reward matrix $\mat{R}$ beforehand where $[\mat{R}]_{j,k,i}=R_{a_i}(s_j,s_k)$, we solve the optimization problem in \eqref{eq:online} using the infinite-horizon MDP. If the ST does not have any prior knowledge about $\mat{T}$ and $\mat{R}$, we employ a reinforcement learning (RL) based algorithm, namely, Q-learning. In this approach, the ST arrives at an optimal policy by taking actions and observing the corresponding rewards. If ST has complete non-causal knowledge about the energy arrivals and channel coefficients up to a finite-horizon (up to $N$ slots), the optimization problem falls within the framework of \textit{offline optimization}. In this case, we first convert \eqref{eq:online}, which is a non-convex MINLP into an equivalent convex MINLP \cite{convex_MINLP}, and then obtain the optimal policy using the GBD algorithm \cite{GBD}.
\section{Online Optimization using MDP}
\label{sec:MDP}
First, we consider the online optimization framework using the infinite-horizon MDP \cite{DP_powell}. Since our goal is to maximize the expected sum-rate of the ST, we employ the \textit{policy iteration} algorithm \cite{DP_powell}. For the online policy, we assume that the ST has complete statistical knowledge of the governing Markov process, \textit{i.e.,} the transition probability matrix $\mat{T}$ and the reward matrix $\mat{R}$ are known in advance. Since the MDP problem in \eqref{eq:online} has finite state and action sets, and the immediate rewards are stationary and bounded, the \textit{policy iteration} algorithm will converge to the optimal policy when $0<\gamma<1$ \cite{MDP_puterman}. The \textit{policy iteration} algorithm obtains the optimal policy in two steps, namely, policy evaluation and policy improvement.

First in the policy evaluation step, the state-value function $V^\pi(s_j),\,\forall s_j\in\mathcal{S}$ under the policy $\pi$ is evaluated (see eq. \eqref{eq:state_value}). Although \eqref{eq:state_value} can be evaluated directly, its computational complexity increases rapidly as the number of states in $\mathcal{S}$ increases. Thus, the \textit{policy iteration} algorithm evaluates the state-value function iteratively \cite{RL_Sutton_book}. Given a policy $\pi$, transition probability matrix $\mat{T}$, and reward matrix $\mat{R}$, the state-value function can be estimated as
\begin{align}
V_l^\pi(s_j)=&C_{s_j}^{s_k}(\pi(s_j))+\gamma\sum_{s_k\in\mathcal{S}} \mathbb{P}_{\pi(s_j)}(s_j,s_k)V_{l-1}^\pi(s_k),\nonumber\\
&\qquad\qquad\qquad\qquad\qquad\qquad\quad \forall s_j\in\mathcal{S},\label{eq:state_value_estimate}
\end{align}
in the $k$th iteration. As $l\to \infty$, $V_l^\pi(s_j)$ converges to true $V^\pi(s_j)$ \cite{RL_Sutton_book}. Then in the policy improvement step, the \textit{policy iteration} algorithm searches for a better policy $\pi'$ such that $Q^{\pi}(s_j,\pi'(s_j))\geq V^\pi(s_j),\,\forall s_j\in\mathcal{S}$ \cite{DP_Bellman}. A better policy can be obtained by employing the greedy policy to $Q^\pi(s_j,a_i)$ in each state, \textit{i.e.,}
\begin{align}
\pi'(s_j) = \underset{a_i\in\mathcal{A}}{\arg\max}\;\;Q^\pi(s_j,a_i).\label{eq:policy_improvement}
\end{align}
The algorithm initializes with an arbitrary policy $\pi(s_j)$ and the state value function $V^\pi(s_j),\,\forall s_j\in\mathcal{S}$. The algorithm first estimates the state-value function for the given policy $\pi$ using \eqref{eq:state_value_estimate}. Then in each iteration, the algorithm looks for a better policy $\pi'$ using \eqref{eq:policy_improvement} and updates the state-value function to $V^{\pi'}(s_j),\,\forall s_j\in\mathcal{S}$. We conclude the convergence of the
algorithm if the policy improvement step yields same policy in two consecutive iterations. The \textit{policy iteration} algorithm is summarized in Algorithm \ref{algo:relative_value}.

\begin{algorithm}
	\caption{Policy iteration algorithm}
	\label{algo:relative_value}
	\begin{algorithmic}
		\State \textbf{1. Initialization:} Initialize state-value function $V^{\pi}(s_j)$, policy $\pi(s_j)$, $\forall s_j\in\mathcal{S}$ arbitrarily, tolerance level $\delta>0$, transition matrix $\mathbf{T}$ and reward matrix $\mathbf{R}$.
		\State \textbf{2. Policy Evaluation:}
		\State Initialize $\Xi$
		\Repeat
		\For{each $s_j\in\mathcal{S}$}
		\State $u\leftarrow V^{\pi}(s_j)$
		\State $V^{\pi}(s_j)\leftarrow C_{s_j}^{s_k}(\pi(s_j))+\gamma\!\!\!\sum\limits_{s_k\in\mathcal{S}}\!\!\mathbb{P}_{\pi(s_j)}(s_j,s_k)V^\pi(s_k)$
		\State $\Xi\leftarrow\min(\Xi,\left\|u-V^\pi(s_j)\right\|)$
		\EndFor		
		\Until {$\Xi<\delta$}
		\State \textbf{3. Policy Improvement:}
		STOP $\leftarrow0$
		\While{STOP $=0$}
		\For {each $s_j\in\mathcal{S}$}
		\begin{align*}
		g&\leftarrow\pi(s_j)\\
		\pi(s_j)&\leftarrow\underset{a_i\in\mathcal{A}}{\arg\max}\;\left[C_{s_j}^{s_k}(a_i)+\gamma\sum\limits_{s_k\in\mathcal{S}}\mathbb{P}_{a_i}(s_j,s_k)V^\pi(s_k)\right]
		\end{align*}
		\If{$g=\pi(s_j)$}
		\State STOP $\leftarrow1$
		\State Set $\pi^*=\pi$
		\EndIf
		\EndFor
		\EndWhile\\
		\Return Optimal policy $\pi^*$.
	\end{algorithmic}
\end{algorithm}
\section{Learning Theoretic Approach}
\label{sec:learning}
Next, we assume that the ST lacks the knowledge about the underlying Markov process, \textit{i.e.,} the transition probability matrix $\mathbf{T}$ and the reward matrix $\mathbf{R}$ are unknown. In this case, we employ a reinforcement learning algorithm called the Q-learning to obtain the harvest-or-transmit policy \cite{DP_powell}. In the Q-learning algorithm, an agent (or user) first takes an action $a_l\in\mathcal{A}$ in a state $S_n\in\mathcal{S}$ and then observes the next state $S_{n+1}$ and receives an instantaneous reward $R_{a_l}(S_n,S_{n+1})$. Since the agent does not know $S_{n+1}$ before taking the action $a_l$, it does not know $R_{a_l}(S_n,S_{n+1})$ either.

We observe from \eqref{eq:action_value} that $Q^{\pi}(S_n,a_l)$ for the current action-state pair contains the immediate expected reward and $V^\pi(S_{n+1})$, which is the state-value function of the next state. Thus $Q^{\pi}(S_n,a_l)$ includes all the long term consequences of taking an action $a_l$ in the state $S_n$ while following a policy $\pi$. Therefore an optimal action can be taken by looking only at $Q^{\pi^*}(s_j,a_l)$, and choosing the action that yields highest expected reward without even knowing the transition probability $\mathbb{P}_{a_l}(S_n,S_{n+1})$ and/or the immediate reward $R_{a_l}(S_n,S_{n+1})$. The Q-learning algorithm obtains the optimal policy by estimating the $Q^{\pi^*}(s_j,a_l)$ recursively. 

The algorithm initializes with an arbitrary $Q_0(s_j,a_l)$ and a state $S_n\in\mathcal{S}$. In each iteration, the agent first takes an action $a_l\in\mathcal{A}$ and observes the next state $S_{n+1}$ and the immediate reward $R_{a_l}(S_n,S_{n+1})$. Then the agent updates the $Q_i(s_j,a_l)$ using the Robbins-Monro stochastic approximation \cite{robbins_monro} as follows
\begin{align*}
Q_i(s_j,a_l) = &(1-\alpha_i)Q_{i-1}(s_j,a_l)\\
&+\alpha_i\left[R_{a_l}(s_j,s_k)+\gamma\max_{a_l\in\mathcal{A}}Q_{i-1}(s_j,a_l)\right],
\end{align*}
where $\alpha_i$ is the learning parameter in the $i$th iteration such that
\begin{itemize}
	\item $0<\alpha_i<1$.
	\item $\sum_{i=1}^\infty \alpha_i=\infty$: To overcome any initial condition.
	\item $\sum_{i=1}^\infty \alpha^2_i<\infty$: To guarantee convergence.
\end{itemize}
If learning rate $\alpha_i$ satisfies the conditions given above, discount factor $\gamma$ satisfies $0<\gamma<1$, and all actions are performed with some probability, the $Q_i(s_j,a_l)$ converges to $Q^{\pi^*}(s_j,a_l)$ with probability 1 as $i\rightarrow\infty$ \cite{learning_Watkins}.

To explore all possible actions to guarantee convergence, we employ the \textit{$\epsilon$-greedy action-selection} method, where with probability $\epsilon$, the algorithm takes a random action, and with probability $1-\epsilon$, the algorithm follows a greedy policy. The convergence rate of the Q-learning algorithm depends on the learning rate $\alpha_i$ and decreases with increase in the number of slots $N$ and the number of learning iterations $N_L$. A more detailed discussion on the convergence of the Q-learning algorithm is presented in \cite{learning_rate_dar}. The Q-learning algorithm is summarized in Algorithm \ref{algo:q_learning}.
\begin{algorithm}
	\caption{Q-learning algorithm}
	\label{algo:q_learning}
	\begin{algorithmic}
		\State \textbf{Initialization:} Initialize $Q_0(s_j,a_l)$ for all $s_j\in\mathcal{S}$, $a_l\in\mathcal{A}$ arbitrarily, $N_L$ and $\epsilon$.
		\State Set STOP $\leftarrow1$ and $i\leftarrow1$
		\State observe starting state $s_j\leftarrow S_0$
		\While{STOP $\neq0$}
		\State Select an action $a_i\in\mathcal{A}$ following $\epsilon$-greedy action selection method
		\State Perform action $a_l\leftarrow a_i$
		\State Observe the next state $s_k\leftarrow S_{i+1}$
		\State Receive the immediate reward $R_{a_l}(s_j,s_k)$
		\State Update
		\begin{align*}
			Q_i(s_j,a_l) \leftarrow& (1-\alpha_i)Q_{i-1}(s_j,a_l)+\alpha_i[R_{a_l}(s_j,s_k)+\\ &\gamma\max_{a_l\in\mathcal{A}}Q_{i-1}(s_j,a_l)],\;\forall s_j\in\mathcal{S},\; a_l\in\mathcal{A}
		\end{align*}
		\State Update current state $s_j\leftarrow s_k$
		\If{$i=N_L$}
		\State STOP $\leftarrow0$
		\EndIf
		\State Update $i\leftarrow i+1$
		\EndWhile 
		\State Set $a^{N_L}(s_j) \leftarrow \underset{a_l\in\mathcal{A}}{\arg\max}\left[Q_{N_L}^{\pi}(s_j,a_l)\right],\;\forall s_j\in\mathcal{S}$ \\
		\Return Policy $\pi^*=\{a^{N_L}(s_1),a^{N_L}(s_2),\ldots,a^{N_L}(s_{N_S})\}$
	\end{algorithmic}
\end{algorithm}

\section{Offline Optimization}
\label{sec:offline}
In this section, we consider the offline optimization of the problem in \eqref{eq:online_orig}, where we assume that channel coefficients and energy arrival amounts are known non-causally at the ST up to $N$ slots. The offline optimization framework can be useful in situations where the underlying stochastic processes governing the system can be estimated accurately. The solution of the offline optimization problem gives an upper bound on the sum-rate achieved by the online and reinforcement learning based policies and can be used to gain useful insights about the optimal harvest-or-transmit policy.

In practice, one can estimate/predict the channel coefficients over a finite number of slots using any channel estimation/prediction technique \cite{channel_estimation_survey_2,fading_prediction,pilot_based_prediction_1,channel_estimation_survey}. Under the offline optimization framework, the optimization problem in \eqref{eq:online_orig} can be rewritten as
\begingroup
\allowdisplaybreaks
\begin{subequations}
\label{eq:offline_nc}
\begin{align}
\hspace{-2mm}\max_{\left\{I_{\rm H}^{[i]},P_s^{[i]}\right\}_{i=0}^N}\quad&\sum_{i=1}^N \gamma^i\left(1-I_{\rm H}^{[i]}\right)\log_2\left(1+\frac{h_{ss}^{[i]} P_s^{[i]}}{\sigma_n^2+h_{ps}^{[i]}P_p}\right)\label{eq:offline_nc_obj}\\
\t{s.t.\hspace{8mm}} & \sum_{j=1}^i \left(1-I^{[j]}_{\rm H}\right) P_s^{[j]}\leq B_0+\eta \sum_{j=1}^{i}I^{[j]}_{\rm H}E^{[j]}_{\rm H},\nonumber\\ 
&\qquad\qquad\qquad\qquad\qquad i=1,\ldots,N,\label{eq:offline_nc_c1}\\
& \hspace{-3mm}\sum_{j=l}^i \left(1-I_{\rm{H}}^{[j]}\right) P_s^{[j]}\leq B_{\rm max}+\eta\sum_{j=l}^i I_{\rm H}^jE_{\rm H}^j,\nonumber\\
&\qquad\qquad\quad i=1,\ldots,N,\;l=1,\ldots,i,\label{eq:offline_nc_c2}\\
& P_s^{[i]}\leq P_{\rm max},\quad i=1,\ldots,N,\label{eq:offline_nc_c3}\\
& I^{[i]}_{\rm H}\in\mathcal{I},\qquad P_s^{[i]}\geq0,\quad i=1,\ldots,N. \label{eq:offline_nc_c4}
\end{align}
\end{subequations}
\endgroup
The above optimization problem is a non-convex MINLP as the variables $\v{i}_{\rm H}$ and $\v{p}_s$ appear in product form. However, taking advantage of the binary nature of the variable $\v{i}_{\rm H}$, we can decouple the optimization variables and reduce the non-convex MINLP to a convex MINLP \cite{convex_MINLP}.
\begin{figure*}[!h]
	\begin{multline}
	\mathcal{L}(\v{p}_s,\v{\pmb{\mu}},\v{\pmb{\lambda}},\v{\pmb{\nu}},{\pmb{\beta}})=\sum_{i=1}^N\gamma^i\log_2\left(1+\frac{h_{ss}^{[i]}P_s^{[i]}}{\sigma_n^2+h_{ps}^{[i]}P_p}\right)+\sum_{j=1}^N\mu_j\left[P_{\rm max}-P_s^{[j]}\right]+\sum_{j=1}^{N}\lambda_j\left[\left(1-I^{[j]}_{\rm H}\right)\left(B_0+\sum_{j=1}^NE^{[j]}_{\rm H}\right)- P_s^{[j]}\right]\\+ \sum_{j=1}^{N}\nu_j\left[B_0+\eta\sum_{i=1}^jI^{[i]}_{\rm H}E^{[i]}_{\rm H}-\sum_{i=1}^{j} P_s^{[i]}\right]+\sum_{j=1}^{N}\sum_{l=1}^{j}\beta_{j,l}\left[B_{\rm max}+\eta\sum_{i=l}^jI^{[i]}_{\rm H}E^{[i]}_{\rm H}-\sum_{i=l}^{j} P_s^{[i]}\right] .\label{eq:lagrangian}
	\end{multline}
	\hrulefill
\end{figure*}
\subsection*{Convex MINLP}
After some manipulations, the reduced convex MINLP is given as:
\begingroup
\allowdisplaybreaks
\begin{subequations}
\label{eq:offline_c}
\begin{align}
\max_{\left\{I^{[i]}_{\rm H},P_s^{[i]}\right\}_{i=0}^N}\quad&\sum_{i=1}^N\gamma^i \log_2\left(1+\frac{h_{ss}^{[i]} P_s^{[i]}}{\sigma_n^2+h_{ps}^{[i]}P_p}\right)\label{eq:offline_c_obj}\\
\t{s.t.\hspace{8mm}} &  P_s^{[i]}\leq \left(1-I^{[i]}_{\rm H}\right)\left(B_0+\sum_{j=1}^NE^{[j]}_{\rm H}\right),\nonumber\\&
\qquad\qquad\qquad\qquad\qquad i=1,\ldots,N,\label{eq:offline_c_c1}\\
& \sum_{j=1}^i  P_s^{[j]}\leq B_0+\eta \sum_{j=1}^{i}I^{[j]}_{\rm H}E^{[j]}_{\rm H},\nonumber\\
&\qquad\qquad\qquad\qquad \qquad i=1,\ldots,N,\label{eq:offline_c_c2}\\
& \sum_{j=l}^i P_s^{[j]}\leq B_{\rm max}+\eta\sum_{j=l}^i I_{\rm H}^jE_{\rm H}^j,\nonumber\\
&\qquad\quad\quad i=1,\ldots,N,\;l=1,\ldots,i,\label{eq:offline_c_c3}\\
& P_s^{[i]}\leq P_{\rm max},\quad i=1,\ldots,N, \label{eq:offline_c_c4}\\
& I^{[i]}_{\rm H}\in\C{I},\quad P_s^{[i]}\geq 0,\quad i=1,\ldots,N. \label{eq:offline_c_c5}
\end{align}
\end{subequations}
\endgroup
The equivalence between \eqref{eq:offline_nc} and \eqref{eq:offline_c} can be proved as follows. When $I^{[i]}_{\rm H}=1$ for some $i$, the constraint \eqref{eq:offline_c_c1} yields $P_s^{[i]}\leq0$. This constraint with \eqref{eq:offline_c_c5} results in $P_s^{[i]}=0$. In this case, constraints \eqref{eq:offline_c_c2} and \eqref{eq:offline_c_c3} doesn't affect the constraint set as the right hand side of the inequalities in \eqref{eq:offline_c_c2} and \eqref{eq:offline_c_c3} are positive numbers. On the contrary, when $I^{[i]}_{\rm H}=0$, since the right hand side of the constraint \eqref{eq:offline_c_c1} has a very large value, it has no effect. In this case, the constraints \eqref{eq:offline_c_c2} and \eqref{eq:offline_c_c3} will dominate and represent the energy neutrality constraints in \eqref{eq:offline_nc_c1} and \eqref{eq:offline_nc_c2}, respectively.

Since the objective function in \eqref{eq:offline_c} is convex $\v{p}_s$ and constraint inequalities are linear in $\v{i}_{\rm H}$ and $\v{p}_s$, the problem in \eqref{eq:offline_c} is a convex MINLP \cite{convex_MINLP}. Thus we can employ the GBD algorithm \cite{GBD} to solve it efficiently. Alternatively, a suboptimal low complexity solution can be obtained using particle swarm optimization \cite{pso_eh}. In the next subsection, we obtain the optimal offline harvest-or-transmit policy using the GBD algorithm.

\subsection{Optimal offline harvest-or-transmit policy using the GBD algorithm}
\label{sec:offline_hot_GBD}
To solve the mixed integer linear programs (MILPs) with complicating variables, J.~F.~Benders \cite{Benders} proposed a cutting-plane based method. In these problem, by fixing these variables, the resulting problem becomes more mathematically tractable. Then, cutting-plane approach is used to obtain the optimal value of the complicating variable. A.~M.~Geoffrion \cite{GBD} extended the work of Benders for MINLPs using non-linear duality theory.

The GBD algorithm solves the MINLP by first decomposing the optimization problem into two subproblems, namely, a \textit{primal problem} (by fixing the integer variables), and a \textit{master problem} (by fixing the real variables), and then solving them iteratively. In our problem, $\v{i}_{\rm H}$ is the complicating variable. In each iteration, the fixed value of integer variable $\v{i}_{\rm H}$ obtained from the previous iteration of the master problem is used to solve the primal problem. Then, the solution of the primal problem $\v{p}_s$ and Lagrange multipliers are used to formulate and solve the next iteration of the master problem. This process continues until convergence. The algorithm initializes by solving the primal problem for some arbitrary $\v{i}_{\rm H}$, $\v{i}_{\rm H}^{(0)}$. The primal and the master problem in the $k$th iteration are given as 

\subsubsection{Primal Problem}
\label{sec:primal}
Let us assume that from the $(k-1)$th iteration of the master problem, we obtain a solution $\v{i}^{(k-1)*}_{{\rm H}}$. Then for the $k$th iteration, the primal problem is given as
\begingroup
\allowdisplaybreaks
\begin{subequations}
\label{eq:primal}
\begin{align}
\max_{\v{P}_s}\quad & \sum_{i=1}^N \gamma^i\log_2\left(1+\frac{h_{ss}^{[i]} P_s^{[i]}}{\sigma_n^2+h_{ps}^{[i]}P_p}\right)\label{eq:primal_obj}\\
\t{s.t.\hspace{4mm}} &  P_s^{[i]}\leq \left(1-I^{[i](k-1)*}_{{\rm H}}\right)\left(B_0+\sum_{j=1}^NE^{[j]}_{\rm H}\right),\nonumber\\
&\qquad\qquad\qquad\qquad\qquad\qquad\quad i=1,\ldots,N, \label{eq:primal_c1}\\
& \sum_{j=1}^i P_s^{[j]}\leq B_0+\eta \sum_{j=1}^{i}I^{[j](k-1)*}_{{\rm H}}E^{[j]}_{\rm H},\nonumber\\
&\qquad\qquad\qquad\qquad\qquad\qquad\quad i=1,\ldots,N, \label{eq:primal_c2}\\
& \sum_{j=q}^i P_s^{[j]}\leq B_{\rm max}+\eta\sum_{j=q}^{i} I^{[j](k-1)*}_{{\rm H}}E^{[j]}_{\rm H},\nonumber\\
&\qquad\qquad\qquad\quad i=1,\ldots,N,\;q=1,\ldots,i, \label{eq:primal_c3}\\
& 0\leq P_s^{[i]}\leq P_{\rm max},\qquad\qquad\quad i=1,\ldots,N,\label{eq:primal_c4}
\end{align}
\end{subequations}
\endgroup
with $I^{[i](k-1)*}_{{\rm H}}$ representing the $i$th element of the vector $\v{i}_{\rm H}$.

The optimization problem in \eqref{eq:primal} is a convex in $\v{p}_s$ \cite{cvx_book} and can be solved efficiently using CVX \cite{cvx}. On solving \eqref{eq:primal} in the $k$th iteration, we obtain a solution $\v{p}_s^{(k)*}$, which is then used to formulate the master problem in the $k$th iteration. Note that \eqref{eq:primal} is convex with linear inequality constraints and non-empty constraint set, which ensures zero duality gap. This makes the Karush-Kuhn-Tucker (KKT) stationarity conditions to be necessary and sufficient \cite{cvx_book}. The Lagrangian $\mathcal{L}(\v{p}_s,\v{\pmb{\mu}},\v{\pmb{\lambda}},\v{\pmb{\nu}},{\pmb{\beta}})$ of the primal problem is given in \eqref{eq:lagrangian} at the top of this page. The KKT stationarity conditions are:
\begin{align*}
\Omega_i-\mu^*_i-\lambda^*_{i}-\!\!\left[\sum_{j=i}^{N}\left(\nu^*_j-\sum_{l=1}^j\beta^*_{j,l}\right)\right]&=0,\\
&
\t{for } i=1,\ldots,N,
\end{align*}
where $\Omega_i=\frac{h_{ss}^{[i]}\gamma^i}{\sigma_n^2+h_{ps}^{[i]}P_p+h_{ss}^{[i]}P_s^{{[i]}*}}$, and $\v{\pmb{\lambda}},\v{\pmb{\nu}},{\pmb{\beta}}$, and $\v{\pmb{\mu}}$ are the dual variables for the constraints \eqref{eq:primal_c1}, \eqref{eq:primal_c2}, \eqref{eq:primal_c3}, and \eqref{eq:primal_c4}, respectively. And the complementary slackness conditions are:
\begin{align*}
\mu_i^*\left[P_s^{{[i]}*}-P_{\rm max}\right]=0,&\\
\lambda_i^*\left[ P_s^{{[i]}*}-\left(1-I^{[i]*}_{\rm H}\right)\left(B_0+\sum_{j=1}^NE^{[j]}_{\rm H}\right)\right]=0,&\\
\sum_{j=1}^i\beta_{i,j}^*\left[B_{\rm max}+\eta\sum_{q=j}^i I^{[q]*}_{\rm H}E^{[q]}_{\rm H}-\sum_{q=j}^{i} P_s^{[q]*}\right]=0,&\\
\nu_i^*\left[\sum_{j=1}^{i} P_s^{[j]*}-B_0-\eta\sum_{j=1}^i I^{[j]*}_{\rm H}E^{[j]}_{\rm H}\right]=0,&\\  i=1,\ldots,&N.
\end{align*}
The dual variables associated with non-negativity constraints can be neglected for mathematical ease, but can be included later using the $\max\{\cdot,0\}$ projection. The optimal transmit power in the $k$th iteration can be obtained using the KKT conditions as
\begin{align}
P_s^{[i](l)*}=\left[\frac{\gamma^i}{\zeta_i}-\frac{\sigma_n^2}{h_{ss}^{[i]}}-\frac{h_{ps}^{[i]}P_p}{h_{ss}^{[i]}}\right]^{+(k)}, & \;i=1,\ldots,N,\label{eq:sec_power_closed}
\end{align}
where
\begin{align*}
\zeta_i &=\lambda^{(l)*}_{i}\!+\mu^{(l)*}_i+\left[\sum_{j=i}^{N}\left(\nu^{(l)*}_j-\sum_{l=1}^j\beta^{(k)*}_{j,l}\right)\right],\\&\qquad\qquad\qquad\qquad\qquad\qquad\qquad i=1,\ldots,N,
\end{align*}
and $[x]^+=\max\{0,x\}$. To obtain $P_s^{[i](k)*}$ in the $k$th iteration using \eqref{eq:sec_power_closed}, we need the optimal values of Lagrange multipliers $\v{\pmb{\lambda}}$, $\v{\pmb{\mu}}$, $\v{\pmb{\nu}}$, and $\pmb{\beta}$, which we can obtain using either CVX \cite{cvx} or iterative dual-descent method \cite{dual_descent}. The master problem for the $k$th iteration is explained in the next subsection.
\subsubsection{Master Problem}
\label{sec:master}
Following two manipulations \cite{GBD} yield the master problem:
\begin{enumerate}
	\item Projecting \eqref{eq:offline_c} onto $\v{i}_{\rm H}$ space as
	\begin{align*}
	\max_{\v{i}_{\rm H}\in\mathcal{I}}\quad g(\v{i}_{\rm H}),
	\end{align*}
where 
\begin{align*}
g(\v{i}_{\rm H})=\left\{\begin{array}{ll}
\sup\limits_{\v{p}_s} & \sum\limits_{i=1}^N \gamma^i\log_2\left(1+\frac{h_{ss}^{[i]} P_s^{[i]}}{\sigma_n^2+h_{ps}^{[i]}P_p}\right)\\
\text{s.t.} & \eqref{eq:offline_c_c1}-\eqref{eq:offline_c_c5},\; P_s^{[i]}\geq 0,\; \forall i.
\end{array}
\right.
\end{align*}
The $g(\v{i}_{\rm H})$ is the primal problem discussed in Section \ref{sec:primal}.
\item Invoke the dual representation of $g$.	
\end{enumerate}
The master problem for the $k$th iteration can be obtained using the above mentioned manipulations as \cite{GBD}:
\begingroup
\allowdisplaybreaks
\begin{subequations}
\label{eq:master}
\begin{align}
\max_{g(\v{i}_{\rm H}),t\geq 0}\quad & t \label{eq:master_obj}\\
\text{s.t.}\hspace{3mm} \quad & t\leq \mathcal{L}\left(\v{p}_s^{(m)*},\bar{\pmb{\mu}}^{(m)*},\bar{\pmb{\lambda}}^{(m)*},\bar{\pmb{\nu}}^{(m)*},{\pmb{\beta}}^{(m)*}\right),\nonumber\\&
\qquad\qquad\qquad \qquad\quad m\in\{1,2,\ldots,k\}\label{eq:master_c1}\\
& g(\v{i}_{\rm H})\in\mathcal{I}. \label{eq:master_c2}
\end{align}
\end{subequations}
\endgroup
The optimization problem in \eqref{eq:master} is an MILP of $t$ and $\v{i}_{\rm H}$ and hence, can be solved optimally using MOSEK \cite{mosek}.

\textbf{Generalized Benders decomposition algorithm:} In the $k$th iteration, the master problem gives an upper bound $t^{*(k)}$ to the solution of the original problem\eqref{eq:offline_c}. Also, in each iteration, one additional constraint (\ref{eq:master_c1}) is added to the master problem. Hence with the number of iterations, this $t^{*(k)}$ is non-increasing. 

The optima of the primal problem $\v{p}_s$ lower bounds to the optimum of the original problem \eqref{eq:offline_c} as it gives a solution for fixed $\v{i}_{\rm H}$, \textit{i.e.,} $\v{i}^{(k-1)*}_{{\rm H}}$. The lower bound of the current iteration is set equal to the maximum of the lower bounds up to current iteration. Hence the lower bound obtained by solving the primal problem is non-decreasing with the number of iterations. For the $k$th iteration, we denote the lower and the upper bounds by Lower$_{(k)}$ and Upper$_{(k)}$, respectively.

The $k$th iteration of \eqref{eq:primal} is solved for the solution of \eqref{eq:master} obtained in the $(k-1)$th iteration. Then, the solution of \eqref{eq:master} in the $k$th iteration is obtained using the solution of the \eqref{eq:primal} obtained in the $k$th iteration. This process continues and since with number of iterations, the solution of the primal (master) problem is non-decreasing (non-increasing), the optima is obtained in finite number of steps \cite{GBD}. The Algorithm \ref{algo:GBD} summarizes the GBD algorithm.
\begin{algorithm}
	\caption{GBD algorithm}
	\label{algo:GBD}
	\begin{algorithmic}
		\State \textbf{Initialization:} Initialize $\v{i}^{(0)}_{{\rm H}}$ randomly, convergence threshold $\Gamma$. Set $\mathcal{B}\leftarrow\emptyset$ and $k\leftarrow1$.
		\State Set $\text{GBD Converge}\leftarrow0$
		\While{$\text{GBD Converge}\neq1$}
		\State Solve \eqref{eq:primal} and obtain
		\{$\v{p}_s^*,\v{\pmb{\mu}}^*,\v{\pmb{\lambda}}^*,\v{\pmb{\nu}}^*,{\pmb{\beta}}^*$\} and Lower$_{(k)}$
		\State $\mathcal{B}\leftarrow \mathcal{B}\cup\{k\}$
		\State solve (\ref{eq:master}) and obtain $\v{i}^{(k)*}_{{\rm H}}$ and the Upper$_{(k)}$.
		\If{$\vert$Upper$_{(k)}$-Lower$_{(k)}\vert\leq\Gamma$}
		\State $\text{GBD Converge}\leftarrow1$
		\EndIf
		\State Set $k\leftarrow k+1$
		\EndWhile \\
		\Return $\v{p}_s$ and $\v{i}_{\rm H}$
	\end{algorithmic}
\end{algorithm}
\begin{theorem*}
	For MINLP \eqref{eq:offline_c}, the $\Gamma$-optimal convergence of the GBD algorithm within finite number of iterations holds for any $\Gamma\geq0$.
\end{theorem*}
\begin{IEEEproof}
A $\Gamma$-optimal solution is obtained by the GBD algorithm if $\left|\text{Upper}_{(k)}-\text{Lower}_{(k)}\right|\leq \Gamma$, for $\Gamma\geq 0$ with $k$ being the iteration number \cite{GBD,MINLP_floudas}. Let $\mathcal{P_S}\subseteq \mathbb{R}_+^{N}$ and $\mathcal{I_H}=\{0,1\}^N$ be the sets such that
\begin{align*}
\mathcal{P_S}&=\{\v{p}_s:P_s^{[i]}\geq0,\,\forall i\} \text{ and }\\
\mathcal{I_H}&=\left\{\v{i}_{\rm H}:I^{[i]}_{\rm H}\in\mathcal{I}=\{0,1\},\, \forall i\right\},
\end{align*}
and $\v{f}(\v{p}_s,\v{i}_{\rm H}):\mathcal{P_S\times I_H}\mapsto \mathcal{X}\subseteq \mathbb{R}^p$ is a vector of functions $f_i(\cdot,\cdot)$ such that $\v{f}(\v{p}_s,\v{i}_{\rm H})\preceq\v{0}$ represent the constraints \eqref{eq:offline_c_c1}-\eqref{eq:offline_c_c4}, and $p$ represents the number of inequality constraints.

Note that in the optimization problem \eqref{eq:offline_c}, the set $\mathcal{P_S}$ is nonempty and convex, and inequalities are affine for each fixed $\v{i}_{\rm H}\in\mathcal{I_H}$. Also, functions $f_i$'s are continuous for each fixed $\v{i}_{\rm H}$. In addition, for fixed $\v{i}_{\rm H}$, the problem \eqref{eq:offline_c} is convex in $\v{p}_s\in\mathcal{P_S}$, and has an optimal solution $\v{p}_s^*$ and dual variables $(\v{\pmb{\mu}}^*,\v{\pmb{\lambda}}^*,\v{\pmb{\nu}}^*,{\pmb{\beta}}^*)$ for linear inequalities \cite{cvx_book}. Thus, following the steps presented in \cite{GBD}, the convergence can be proven for the GBD algorithm for any $\Gamma\geq0$.
\end{IEEEproof}
\section{Myopic Policy}
\label{sec:myopic}
Now we consider the myopic policy, where the ST consumes all the energy it has harvested in the same slot. We adapt the myopic policy proposed in \cite{underlay_3} for comparison which follows time-sharing between the harvesting and transmission phases. To make the policy proposed in \cite{underlay_3} compatible with our assumptions for a fair comparison, we reformulate the optimization problem of obtaining optimal time-sharing.

Let $\alpha_i$ denote the time-sharing parameter such that $(1-\alpha_i)$ fraction is used for energy harvesting, and $\alpha_i$ fraction is used for data transmission. Then, the total energy harvested by ST in the $i$th slot is given as
\begin{align*}
E_{\rm harvested}^{[i]} = (1-\alpha_i)\cdot \eta\cdot E_{\rm H}^{[i]},
\end{align*}
and the transmit power of ST in $i$ th slot is given as
\begin{align*}
P_s^{[i]} = \frac{(1-\alpha_i)}{\alpha_i}\cdot\eta\cdot E_{\rm H}^{[i]}.
\end{align*}
Since the transmit power becomes a function of time-sharing parameter for myopic policy, the optimization problem of designing optimal transmission policy reduces to obtaining optimal time-sharing parameter $\bar{\boldsymbol{\alpha}}$. The optimization problem is given as
\begin{subequations}
	\begin{align}
	\max_{\bar{\boldsymbol{\alpha}}}\quad & \sum_{i=1}^N\gamma^i\alpha_i\log_2\left(1+\frac{(1-\alpha_i)}{\alpha_i}\cdot\frac{h_{ss}^{[i]}\eta E_{\rm H}^{[i]}}{\sigma_n^2+h_{ps}^{[i]}P_p}\right) \label{eq:myopic_obj}\\
	\st & (1-\alpha_i)\eta E_H^{[i]} \leq \alpha_i P_{\rm max}, \qquad i=1,\ldots,N,\label{eq:myopic_c1}\\
	& \bar{\mathbf{0}}\preceq \bar{\boldsymbol{\alpha}} \preceq \bar{\mathbf{1}}. \label{eq:myopic_c2}
	\end{align}
\end{subequations}
The above problem is a convex optimization problem since the objective function is a sum of negative relative entropies, $D(p_i\| q_i)$ where $p_i = \gamma^i\alpha_i$ and $q_i = \gamma^i\left(\alpha_i+(1-\alpha_i)\cdot\frac{\eta h_{ss}^{[i]}E_{\rm H}^{[i]}}{\sigma_n^2+h_{ps}^{[i]}P_p}\right)$, and constraints are linear inequalities. Therefore, this problem can be solved optimally using CVX \cite{cvx}.

\section{Results and Discussions}
\label{sec:result}
This section presents the simulation results for offline, online and learning-theoretic harvest-or-transmit policies discussed in previous sections. We assume that the harvested energy can take values from the set $\mathcal{E}=\{0.2,0.4\}$ mJ, and channel power gains $h_{pp}^{[i]}, h_{ps}^{[i]}, h_{sp}^{[i]}$, and $h_{ss}^{[i]}$ can take values from the set $\mathcal{H} = \{0.2, 0.4\}\times 10^{-6}$ as in \cite{underlay_3}. The transmit power of the PT, $P_p$ is assumed to be $2$ mW in all the slots, and the noise power at the SR, $\sigma_n^2$ is assumed to be $-90$ dBm. Unless otherwise stated, we assume the battery capacity at the ST to be $10$ mJ. The worst-case interference constraint at the PR is $P_{\rm int}=0.4$ nW, which results in the maximum allowed transmit power of the ST, $P_{\rm max}$ to be 1 mW. To obtain the online policy, we consider uniform state transition probability, \textit{i.e.}, the probability of transition from any state $s_j$ to any state $s_k$ is $\mathbb{P}(s_j,s_k)=0.125$. For the online and learning-theoretic policies, the action space for the transmit power of the ST is discrete and can take values from the set $\{0,0.2,\cdots, P_{\rm max}\}$ mW.

\subsection{Optimal Action-Selection Probability $\epsilon$}
\begin{figure}[!ht]
		\centering
	\includegraphics[width=0.95\linewidth]{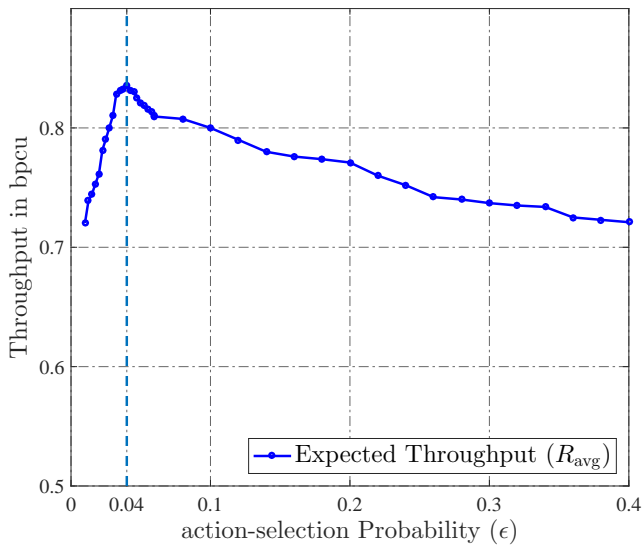}
	\caption{Expected throughput ($R_{\rm avg}$) with respect to action-selection probability ($\epsilon$).}
	\label{fig:ch5_throughput_vs_act_select}
\end{figure}

The Fig. \ref{fig:ch5_throughput_vs_act_select} shows the average achievable throughput of the ST, $R_{\rm avg}$ versus the action selection probability, $\epsilon$ for $N_L=40$. In the figure, observe that as $\epsilon$ increases, the expected throughput first increases and then decreases. This because, initially as $\epsilon$ increases, the ST starts exploring more possible actions by taking random actions more often, which improves its achievable throughput. However, if we further increase $\epsilon$, the ST will increase the number of random actions, which reduces its achievable throughput. Using cross validation, we obtained the optimal action-selection probability $\epsilon=0.04$ which maximizes the achievable throughput of the ST for given number of learning iterations.

\subsection{Comparison of Proposed Policies}
\label{subsec:ch5_comparison}
\begin{figure}[!ht]
	\centering
	\includegraphics[width=0.95\linewidth]{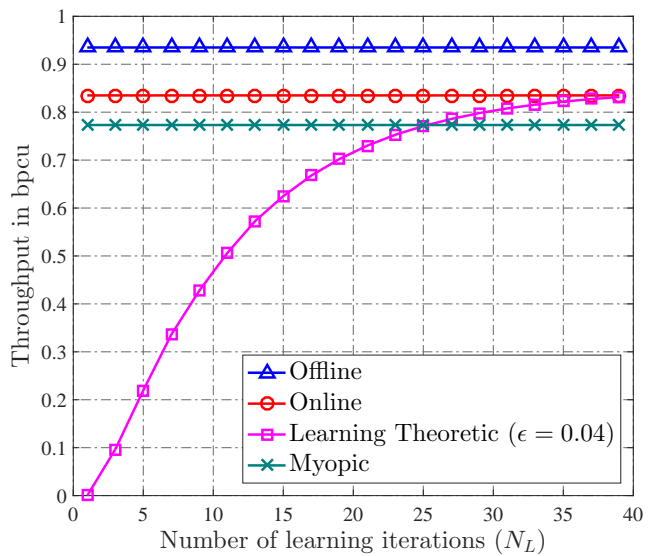}
	\caption{Expected throughput ($R_{\rm avg}$) with respect to number of learning iterations ($N_L$).}
	\label{fig:ch5_throughput_vs_iteration}
\end{figure}
The Fig. \ref{fig:ch5_throughput_vs_iteration} shows the average achievable throughput of the ST, $R_{\rm avg}$ versus the number of learning iterations, $N_L$ under all harvest-or-transmit policies along with myopic policy proposed in \cite{underlay_3}. In the figure, observe that the offline policy outperforms others and acts as a benchmark for the online and learning-theoretic harvest-or-transmit policies. The learning theoretic policy, on the other hand, starts by performing a random action with probability $\epsilon=0.04$ and follows the greedy policy with probability $\epsilon=0.96$. In each iteration, it updates the Q-function depending on the next state and received reward. Therefore, the average achievable throughput increases with $N_L$. Also, all proposed policies outperform the myopic policy proposed in \cite{underlay_3} because, the myopic policy only aims to maximize the instantaneous which might not be optimal in fading conditions over a large number of slots.

\subsection{Effects of action-selection probability $\epsilon$}
\label{subsec:ch5_effect_of_epsilon}

\begin{figure}[!ht]
	\centering\includegraphics[width=0.95\linewidth]{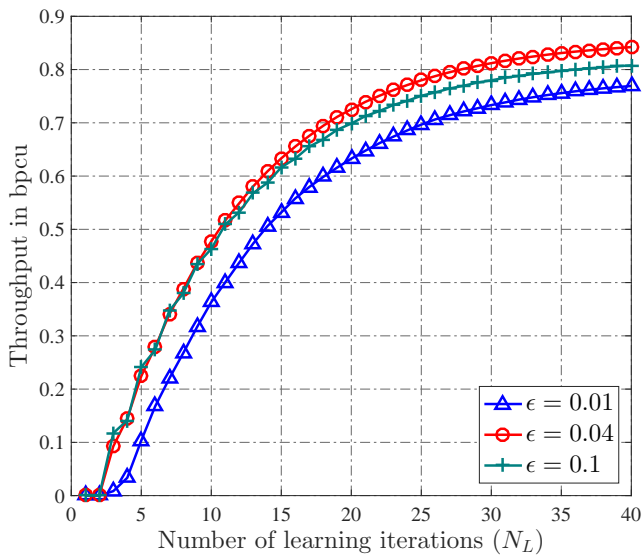}
	\caption{Expected throughput ($R_{\rm avg}$) with respect to number of learning iterations ($N_L$) for different action-selection probability ($\epsilon$).}
	\label{fig:ch5_throughput_vs_iteration_diff_act_select}
\end{figure}

Fig. \ref{fig:ch5_throughput_vs_iteration_diff_act_select} shows the effects of the probability of action-selection, $\epsilon$ on the learning theoretic policy for a single channel realization. Observe that when $\epsilon=0.01$, the learning rate of the Q-learning algorithm is less. This is because when $\epsilon$ is low, the algorithm focuses on the greedy policy by taking only the known actions (with probability $0.99$) and does not explore the new possible actions. In this case, the algorithm converges to a sub-optimal solution. If we increase the $\epsilon$, the algorithm starts exploring the new possible actions more often, and therefore, the learning rate increases. However, if we further increase the $\epsilon$ (say $\epsilon=0.1$), the algorithm takes more random actions and does not follow the greedy policy. This increases the learning rate initially, but since the algorithm is not following the greedy policy, after exploring all possible actions, its learning rate reduces, and the algorithm converges to a suboptimal solution. In the figure, observe that the optimal action-selection $\epsilon=0.04$ yields the best performance.
\begin{figure}[!ht]
	\centering\includegraphics[width=0.95\linewidth]{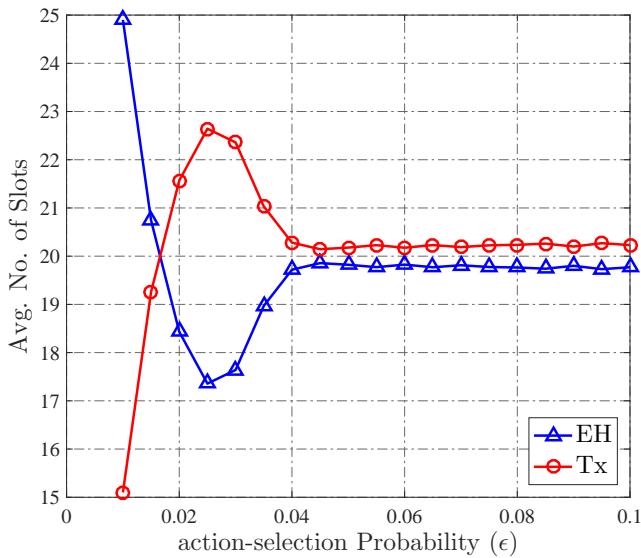}
	\caption{Average number of harvesting and transmission slots with respect to action-selection probability ($\epsilon$).}
	\label{fig:ch5_slots_vs_action_selection}
\end{figure}


\begin{figure}[!ht]
	\centering\includegraphics[width=0.95\linewidth]{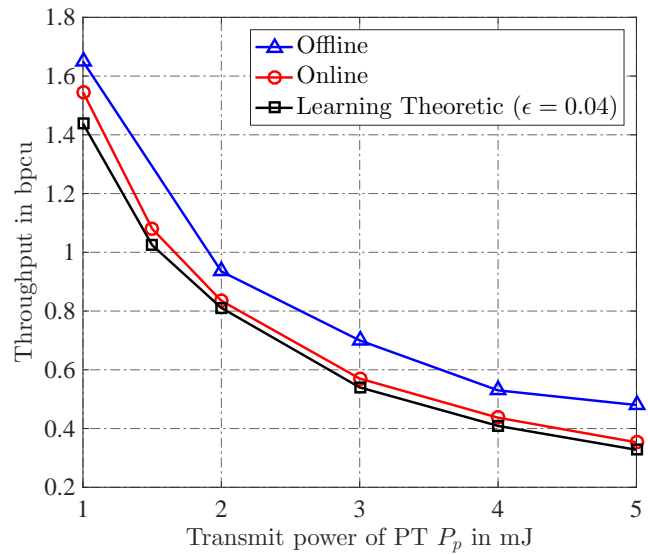}
	\caption{Expected throughput ($R_{\rm avg}$) with respect to transmit power of primary user ($P_p$).}
	\label{fig:ch5_throughput_vs_P_p}
\end{figure}

Fig. \ref{fig:ch5_slots_vs_action_selection} shows the effects of action-selection probability $(\epsilon)$ on the number energy harvesting and transmission slots under the learning-theoretic policy for $N_L=40$ averaged over 2000 state realizations. In the figure, observe that as $\epsilon$ increases, the number of harvesting (transmission) slot decreases (increases) initially. This is because, for small $\epsilon$, the algorithm follows a greedy policy and does not explore new possible actions. Thus, it continues to choose ``harvesting'' over ``transmission'', which in turn reduces the achievable throughput as shown in Fig. \ref{fig:ch5_throughput_vs_iteration_diff_act_select}. As we increase $\epsilon$, the algorithm starts exploring new possible actions and therefore the number of harvesting (transmission) slots decreases (increases). However, if we further increase $\epsilon$, the algorithm takes random actions more often due to which, the number of harvesting slots starts increasing. In the figure, for the optimal $\epsilon$ ($\epsilon=0.04$), the number of harvesting and transmission slots is approximately the same.

\subsection{Effects of Primary's Transmit Power $P_p$}
\label{subsec:ch5_effect_of_p_p}
Fig. \ref{fig:ch5_throughput_vs_P_p} shows the effects of the transmit power of PT on the average achievable throughput. In the figure, observe that the achievable throughput in all the policies decreases as $P_p$ increases. This is because, increasing $P_p$ causes more interference at the SR, which reduces the achievable sum-rate.

\subsection{Effects of Maximum Transmit Power Constraint $P_{\rm max}$}
\label{subsec:effect_of_p_max}
\begin{figure}[!ht]
	\centering\includegraphics[width=0.95\linewidth]{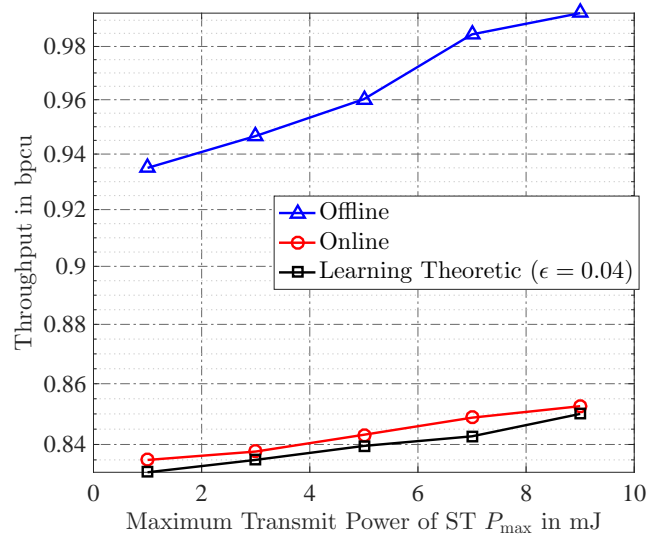}
	\caption{Expected throughput ($R_{\rm avg}$) with respect to maximum transmit power of secondary user ($P_{\rm max}$).}
	\label{fig:ch5_throughput_vs_P_max}
\end{figure}


\begin{figure}[!ht]
	\centering\includegraphics[width=0.95\linewidth]{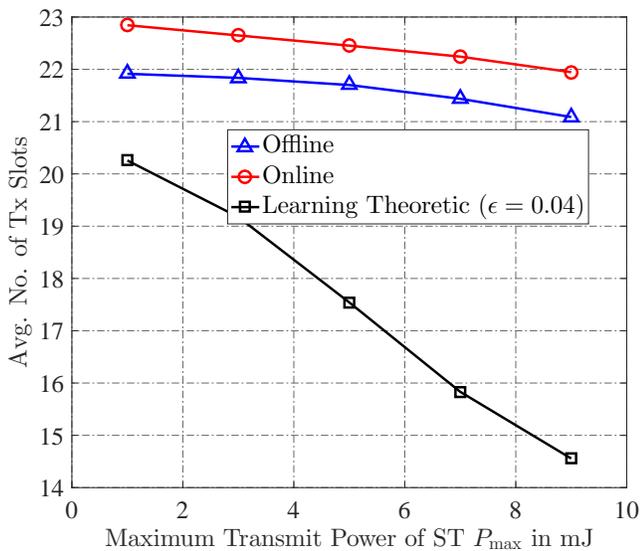}
	\caption{Average number of transmission slots of ST with respect to maximum transmit power of secondary user ($P_{\rm max}$).}
	\label{fig:ch5_slots_vs_P_max}
\end{figure}
\begin{figure}[!ht]
	\centering\includegraphics[width=0.95\linewidth]{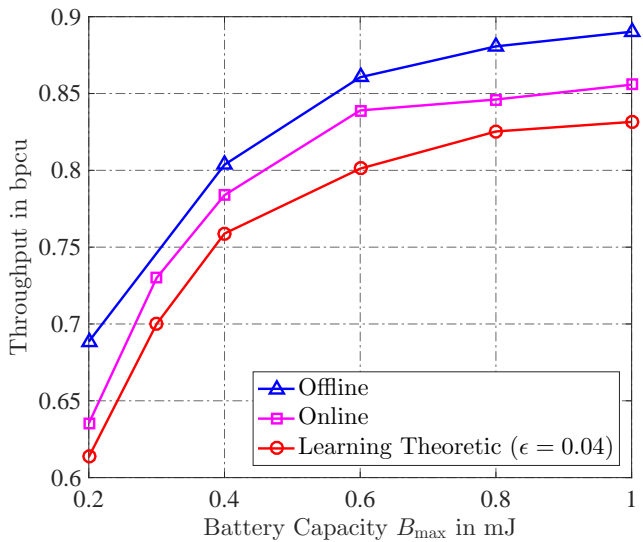}
	\caption{Expected throughput ($R_{\rm avg}$) with respect to battery capacity ($B_{\rm max}$).}
	\label{fig:ch5_throughput_vs_B_max}
\end{figure}
Figs. \ref{fig:ch5_throughput_vs_P_max} and \ref{fig:ch5_slots_vs_P_max} show the effects of $P_{\rm max}$ on the average achievable throughput and the average number of transmission slots of ST, respectively, under all policies. In the Fig. \ref{fig:ch5_throughput_vs_P_max}, observe that the average achievable throughput of ST increases with $P_{\rm max}$ for all policies. This is because increasing $P_{\rm max}$ allows the ST to transmit with a higher power, which increases the throughput. On the other hand, the total number of transmission slots of ST decreases with $P_{\rm max}$ as shown in Fig. \ref{fig:ch5_slots_vs_P_max}. This is because increasing $P_{\rm max}$ allows the ST to transmit with a higher power, which requires higher energy availability. Thus, on increasing $P_{\rm max}$, the ST harvests more amount of time and then consumes the harvested energy in short duration by transmitting with high power.

\subsection{Effects of Battery Capacity $B_{\rm max}$}
\label{subsec:ch5_effect_of_B_max}
Fig. \ref{fig:ch5_throughput_vs_B_max} shows the effect of $B_{\rm max}$ on the expected sum-rate of ST under all policies. The maximum transmit power of the ST is assumed to be 1 mW. In the figure, observe that as the battery capacity increases, the average throughput under all policies increases. This is because, upon increasing the battery size, the ST can store more energy and if channel conditions are good, it can transmit with a higher power. However, if we keep on increasing the battery capacity, the average throughput does not increase any more as the battery capacity becomes large enough to accommodate the harvested energy even if ST does not transmit for multiple slots.

\section{Future Directions}
\label{sec:future}
In this paper, we have assumed sufficiently large amount of backlogged data so that the ST always has some data to send. However for various networks like IoT, this might not be true and the data itself might arrive during the course of communication. This puts an additional constraint on the throughput maximization problem namely ``data causality constraint'' which is similar to the energy causality constraint discussed in Section \ref{subsec:battery_dynamics}. In \cite{EH_scheduling_random_data_1} and \cite{EH_scheduling_random_data_2}, the authors considered random data arrival under offline and online framework, respectively, and obtained optimal transmission policies considering energy and data causality constraints. As a future extension to this work, we can consider a general scenario where we use the concept of MDP to obtain optimal transmission policy considering data and energy queues with finite arrival rates.

\section{Conclusion}
\label{sec:conclusion}
In this paper, we considered an underlay cognitive radio network where a primary-secondary user (PU-SU) transceiver pair operate in a slotted mode. The primary transmitter (PT) has a reliable power supply and transmits with constant power in each slot, whereas the secondary transmitter (ST) chooses an optimal policy of data transmission or energy harvesting depending on the channel conditions and energy availability. The optimal per slot transmit power of the ST is obtained that maximizes its sum-rate. We modeled the fading and energy arrivals as first-order stationary Markov process and obtained the optimal policies under complete, statistical and no knowledge of channel gains and energy arrivals. 

First, we assume that the ST has full statistical knowledge about the underlying Markov process. In this case, we obtained the optimal online harvest-or-transmit policy using \textit{policy iteration} algorithm. Then, we consider a case when ST has no statistical knowledge about the governing Markov process. In this case, we employed the tools from reinforcement learning (RL) and obtained the harvest-or-transmit policy using the \textit{Q-learning} algorithm. Finally, we considered the offline optimization framework where we assumed that the ST has complete non-causal knowledge of channel coefficients and energy arrivals. We formulated the offline optimization problem as a MINLP and obtained the optimal policy using the GBD algorithm.

Finally, we compared the performance of all proposed policies and analyzed the effects of various system parameters on them. We showed that the offline policy yields the best sum-rate and acts as a benchmark policy for online and learning-theoretic policies. The learning theoretic policy, on the other hand, starts from taking random and greedy actions and therefore results in lower sum-rate. However, as the learning progresses, the learning theoretic policy starts performing better and performs close to the online policy asymptotically. For the RL based policy, we obtained an optimal action-selection probability that maximized the ST's expected sum-rate using cross validation. We observed that as the action-selection probability increases, the average number of harvesting (transmission) slots increases (decreases) initially. However, as the action-selection probability increases further, the number of harvesting and transmission slots become approximately the same due to the increased number of random actions. We studied the effects of maximum transmit power constraint on the number of harvesting and transmission slots as well under all the policies, and observed that as we allow the ST to transmit with higher power, it starts harvesting for more number of slots to accumulate more energy. Also, we analyzed the effects of battery capacity on the expected sum-rate and observed that as battery capacity increases, the sum-rate increases due to higher energy availability.

%
%

%
%
%
%
%

\ifCLASSOPTIONcaptionsoff
  \newpage
\fi



%

\bibliographystyle{ieeetr} 
\bibliography{references}

%
%
%

\end{document}